\documentclass[12pt]{article}

\usepackage[margin=1.3in]{geometry}
\usepackage{setspace}
\usepackage{indentfirst}
\usepackage[explicit]{titlesec}
\usepackage{lastpage}
\usepackage[figure,table]{totalcount}
\usepackage{amsmath}
\usepackage{graphicx}
\usepackage{multirow}
\usepackage{caption}
\usepackage[labelformat=simple]{subcaption}
\usepackage[colorlinks=true, allcolors=blue, pdfencoding=auto]{hyperref}
\usepackage{footmisc}

\usepackage{xstring}
\usepackage{etoolbox}
\usepackage{endnotes}

\renewcommand{\thesection}{\Roman{section}}
\renewcommand{\thesubsection}{\thesection.\Alph{subsection}}
\renewcommand{\thesubsubsection}{\thesubsection.\arabic{subsubsection}}
\titleformat{\section}{\bfseries}{\thesection.}{1em}{\MakeUppercase{#1}}
\titleformat{\subsection}{\bfseries}{\thesubsection.}{1em}{#1}
\titleformat{\subsubsection}{\itshape}{\thesubsubsection.}{1em}{#1}

\renewenvironment{abstract}{\pagebreak~\vfill\begin{center}\textbf{Abstract}\end{center}}{\vfill\pagebreak}

\usepackage{graphicx}
\usepackage{amssymb}
\newlength{\mycolumnwidth}
\begin{document}

\begin{titlepage}
    \centering
    \vspace*{2cm}

    {\LARGE\bfseries The Atomic Beam Probe Synthetic Diagnostic and its Application in Fusion Plasmas\par}
    \vspace{1cm}

    {\large
        Mátyás Aradi\textsuperscript{$\ast$,a},
        Dániel I. Réfy\textsuperscript{b},
        Shimpei Futatani\textsuperscript{a},
        Ors Asztalos\textsuperscript{c,b},
        Miklós Berta\textsuperscript{d},
        Pavel Háček\textsuperscript{e},
        Jaroslav Krbec\textsuperscript{e},
        Sándor Zoletnik\textsuperscript{b},
        Gergo I. Pokol\textsuperscript{c,b}
    \par}

    \vspace{0.5cm}

    {\small\
        \textsuperscript{a}\emph{Universitat Politècnica de Catalunya, Barcelona, Spain}\\
        \textsuperscript{b}\emph{HUN-REN Centre for Energy Research, Budapest, Hungary}\\
        \textsuperscript{c}\emph{NTI, Budapest University of Technology and Economics, Budapest, Hungary}\\
        \textsuperscript{d}\emph{Széchenyi István University, Gy\H{o}r, Hungary}\\
        \textsuperscript{e}\emph{Institute of Plasma Physics of the CAS, Prague, Czech Republic}\\        
        \textsuperscript{$\ast$}{email: \href{mailto:matyas.aradi@upc.edu}{matyas.aradi@upc.edu}}
    \par}

    \vfill
    {\large \par}
\end{titlepage}

\begin{abstract}
The Atomic Beam Probe (ABP) is a novel beam diagnostic concept that opens new opportunities in plasma edge measurements due to the sensitivity of the magnetic field and the high temporal resolution. The first ABP was installed and was operating on the COMPASS tokamak. A synthetic diagnostic was implemented with a massively parallel trajectory solver core that runs on graphic cards to support experiments, providing a better understanding of measurements and opening opportunities for future applications. This paper presents the model concept with relevant physical processes and necessary simplifications. The submodules implemented or integrated into the synthetic diagnostic are explained and described, and their scope of validity is highlighted. We demonstrate the utilization of the ABP synthetic diagnostic by comparing the measurements. The synthetic diagnostic was named as \emph{TAIGA Synthetic Diagnostic} (or shortly \emph{TAIGA-SD}).
\end{abstract}

\section{Introduction}
\label{sec:intro}

Alkali Beam Emission Spectroscopy (BES) is an active plasma diagnostics tool for determining plasma density profiles and density fluctuations\cite{zoletnik2018}. High energy -- up to 100~keV -- neutral beam, typically a hydrogen or light alkali beam, is injected into the plasma. The beam atoms are excited due to collisional processes with the plasma particles, and the spontaneous photon emission can be observed by an optical observation system\cite{asztalos2019}.

In the mid-2000s, the idea was raised that the alkali Beam Emission Spectroscopy (BES) system could be extended by detecting singly ionized ions from the injected alkali beam\cite{berta2009}. The principle itself is reminiscent of the Heavy Ion Beam Probe (HIBP), an ion beam that leaves the plasma would be detected outside the plasma, albeit a singly ionized light alkali ion beam. The new diagnostic tool concept was named Atomic Beam Probe (ABP).

In a nutshell, the concept of the Atomic Beam Probe is as follows: a neutral alkali atomic beam is injected perpendicularly into the plasma at the midplane port\cite{berta2013}. The beam particles are ionized, typically in the plasma edge region following the last closed flux surface. The singly ionized ions interact with the magnetic field and are deflected from the midplane\cite{berta2009, berta2013, hacek2018, zoletnik2018}. The trajectories of ions are determined by the magnetic field they pass through, and their measured displacement can be employed to infer changes in the magnetic field. For example, on COMPASS tokamak, the ions were collected with a Faraday-cup matrix detector\cite{zoletnik2018,refy2019}, which allowed sampling on the microsecond timescale. Consequently, fast magnetic field changes on this timescale can be detected.

Similar diagnostics have been developed or upgraded in recent years, so the knowledge gained can be applied to other cases. One such diagnostic is the imaging heavy ion beam probe (i-HIBP) diagnostic on the ASDEX tokamak\cite{birkenmeier2019, birkenmeier2021}, where a heavy neutral, typically rubidium, sometimes cesium, beam is injected into the plasma at an energy between 60 and 72~keV\cite{birkenmeier2019, galdon2024}, and the singly ionized beam is detected by a scintillator detector. Another sister diagnostic is the heavy ion beam probe (HIBP) with sodium, potassium, and cesium single ionized beam is used with an energy up to 100~keV on TUMAN-3M tokamak \cite{askinazi2022} with a secondary ion energy analyzer that allows measuring plasma potential. Another HIBP that has been operating since the 1990s on ISTTOK tokamak with xenon, cesium, and mercury single-ionized beam is injected with an energy between 20 and 25~keV \cite{malaquaias2022}. A new HIBP design was simulated for QUEST tokamak \cite{ido2022} with an injected singly ionized cesium ion beam with an energy of several ten keVs. Other alkali ion beams were investigated in earlier phases of the QUEST-HIBP studies\cite{nifs2012, nifs2013}, which can be relevant for the ABP.

The Atomic Beam Probe was installed on the COMPASS tokamak in Prague as an extension of the Beam Emission Spectroscopy\cite{zoletnik2018}. The construction and installation of the diagnostic were preceded by conceptional design \cite{berta2009} and measurements with a test detector\cite{berta2013}. This development work was supported by a trajectory solver, written as a sequential code, called \emph{ABPIons}\cite{berta2013}, whose shortcomings were discovered early on. For further modeling, orders of magnitude larger numerical performance were required. Therefore, the trajectory calculator has been redesigned and implemented as a GPGPU (general-purpose computing on graphics processing units) code in CUDA C language, which resulted in a performance improvement of three orders of magnitude. The new solver, the \emph{Trajectory simulator of ABP Ions with GPU Acceleration (TAIGA)}, was used to establish the final diagnostics\cite{hacek2018, zoletnik2018}. To aid the improvement of the diagnostics and to design further measurement scenarios, we improved the code and developed a complete synthetic diagnostic \cite{zoletnik2018,refy2019,refy2021}. The trajectory solver is an essential tool for designing the ABP. However, a more sophisticated tool was required to consider all essential physical parameters in the simulation. Therefore, we developed the \emph{TAIGA-SD}\footnote[2]{https://github.com/taiga-project} synthetic diagnostic, which is the subject of the recent paper.

In this article, we introduce the concept of synthetic diagnostics. We then show the initial input conditions, followed by the trajectory simulator, the mathematical and physical models we used, and the synthetic signal processing. Finally, we present how synthetic diagnostics supported the measurements and what the relation is between the synthetic and the measurement data.

\section{The Concept of the Synthetic diagnostic}
\label{sec:concept}

Synthetic diagnostics describes a complex, multi-physics numerical toolset that generates a simulated signal that can be directly compared to an actual, physical measurement \cite{asztalos2019}. This simulation package takes into account all the relevant physical processes. The aim is for the simulation result to be processed in the same way as in the actual physical measurement and for a physical signal to be obtained (electric current in our case). The synthetic signal can then be compared with the corresponding values of the actual measurements.

We have systematically investigated the physical processes and their accuracy using measurement results. Consequently, we determined the physical process submodules of which a synthetic diagnostic should consist. Each process is implemented as a separate unit; sometimes, code from an external source is interfaced. Throughout the paper, we will go through the submodules of the synthetic diagnostic, showing their physical and numerical significance.

\begin{figure}
    \centering
    \includegraphics[width=\mycolumnwidth]{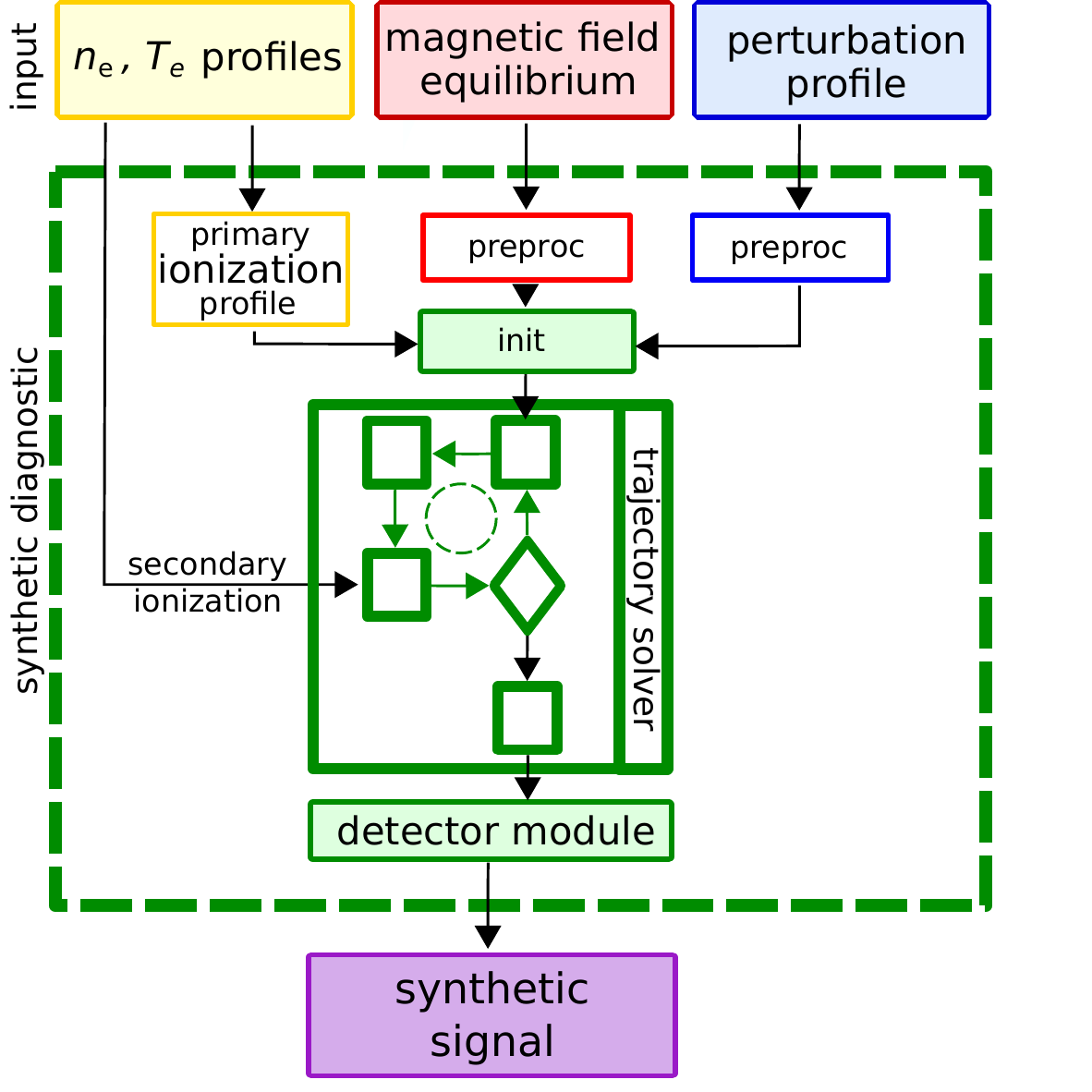}
    \caption{\label{fig:abpsd}Block diagram for the Atomic Beam Probe Synthetic Diagnostic (TAIGA-SD).}
\end{figure}

Figure~\ref{fig:abpsd} shows the block diagram of \emph{TAIGA-SD} synthetic diagnostics, which digitally mimics the Atomic Beam Probe. The submodules are explained in the following sections. In Section~\ref{sec:init} and the green box in the figure, we showcase the synthetic diagnostic initialization, the required input profiles, and the applied physical and numerical models. With the red box, the Subsection~\ref{sec:magnetic} shows the magnetic field initialization and verification. Subsection~\ref{sec:primary}, with a yellow box, explains the primary ionization profile of the physical models and the required experimental profiles behind it. The blue box represents that we can give the function pointer to the synthetic diagnostic in this step, whereas the proper perturbation models are the topic of further studies. The number of singly ionized alkalis depends on the secondary ionization rate, and the model is explained in Subsection~\ref{sec:secondary} as an input for the trajectory solver. The physics and the numerical models behind the \emph{TAIGA} trajectory solver are detailed in Section~\ref{sec:trajectory}. The synthetic signal generation and processing from trajectory details are made in Section~\ref{sec:synthetic}, shown with a purple box. The green boxes highlight all modules executed on the graphic card: these units needed massive parallelization due to high-performance requirements.

\section{Initial boundaries}
\label{sec:init}

For the design of the synthetic diagnostics, we selected the discharge of the COMPASS tokamak that is relevant for our study: discharge \#17178 at time 1097~ms. In this measurement, a Li-BES experiment was set up in the relevant range and tested the ABP system on a reduced-resolution detector unit \cite{zoletnik2018}. In the following, we refer to this discharge as the reference discharge, providing the magnetic field, electron density, and temperature profiles.

\subsection{Equilibrium magnetic field}
\label{sec:magnetic}

The ABP can be utilized to measure rapid fluctuations; nonetheless, the course time-resolved reconstructions that EFIT\cite{appel2006} can provide are also relevant for diagnostic development and the validation of synthetic diagnostics. Furthermore, the EFIT reconstruction data can be employed as reference geometry for rapid fluctuation modeling.

The EFIT++\cite{appel2006} solves the Grad--Shafranov equation and results in the flux coordinates\cite{appel2006}. The tool is applied to many fusion devices and is also available at COMPASS tokamak\cite{urban2015}. The spatial accuracy of the reconstruction is about 1~cm on the COMPASS tokamak\cite{urban2015, jirakova2019}. The magnetic flux coordinates have a time resolution of one millisecond and are averaged over time. From the flux coordinates, we can obtain the components of the equilibrium magnetic field by numerical derivation.

\begin{figure}
    \centering
    \includegraphics[width=0.95\mycolumnwidth]{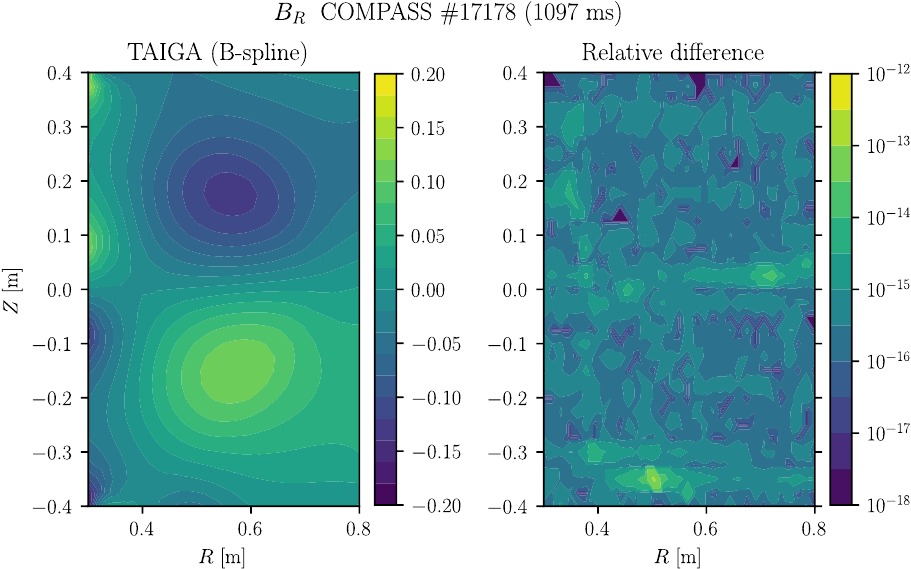}
    \includegraphics[width=0.95\mycolumnwidth]{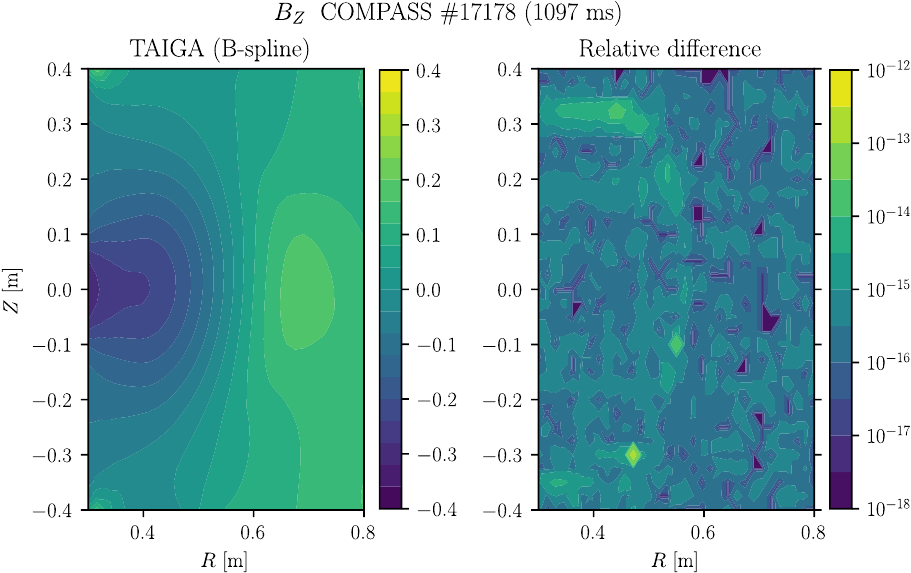}
    \includegraphics[width=0.95\mycolumnwidth]{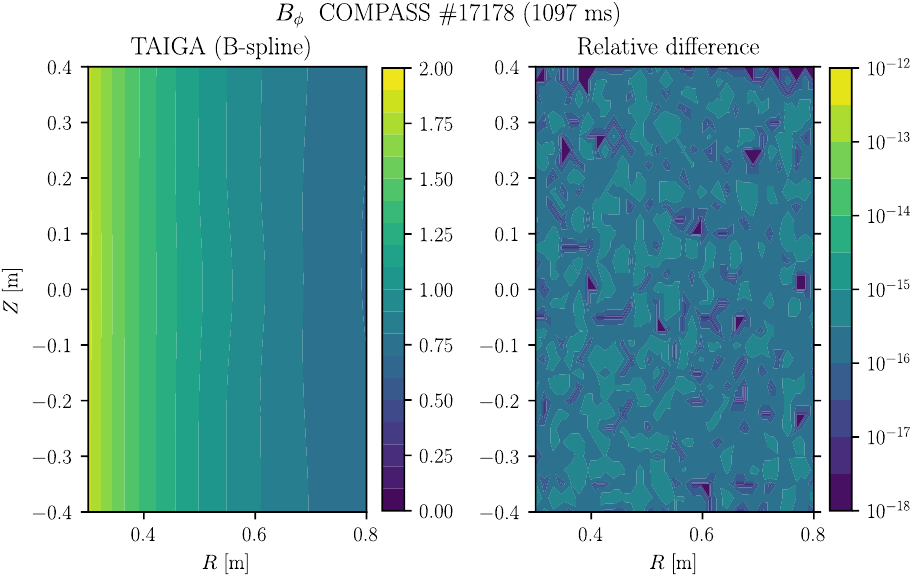}
    \caption{\label{fig:brad}Comparison of TAIGA (on left columns) and the difference comparing to pyCDB\cite{urban2015} (on right columns) magnetic field reconstruction.}
\end{figure}

A bicubic B-spline was fitted to the grid points using a custom Python interface, and the spline coefficients were extracted. The numerical trajectory solver requires the knowledge of the magnetic field at the current position of the ion in every iteration. The magnetic field is calculated on a spatial grid, so interpolation is required in the intermittent positions. The coefficients are calculated and stored during the initialization for better numerical performance. Our magnetic field reconstruction was benchmarked to COMPASS' routine (pyCDB) \cite{urban2015}. All magnetic field components show good agreement, as presented in~Figure~\ref{fig:brad}.

Our magnetic field reconstruction submodule gives a magnetic field on a two-dimensional poloidal cross-section, and we assume toroidal symmetry when using our synthetic diagnostic in tokamaks.

\subsection{Primary ionization profile}
\label{sec:primary}

Knowing the magnetic field, we can calculate the trajectories of the ions, whereas this requires some initial parameters, such as the coordinates of ionization. To mimic the Atomic Beam Probe, we need to know where the neutral beam particles are ionized along the beam. The ionization profile calculations require electron temperature and density profiles and the parameters of the atomic beam, such as geometry, energy, and species. The profiles are typically reconstructed from Thomson scattering experiments.

\begin{figure}
    \centering
    \includegraphics[width=0.45\textwidth]{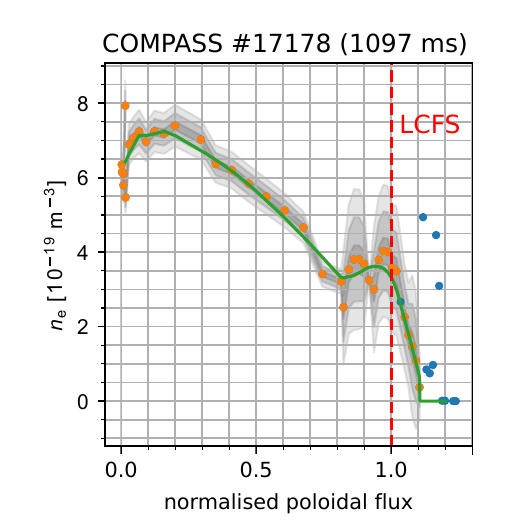}
    \includegraphics[width=0.45\textwidth]{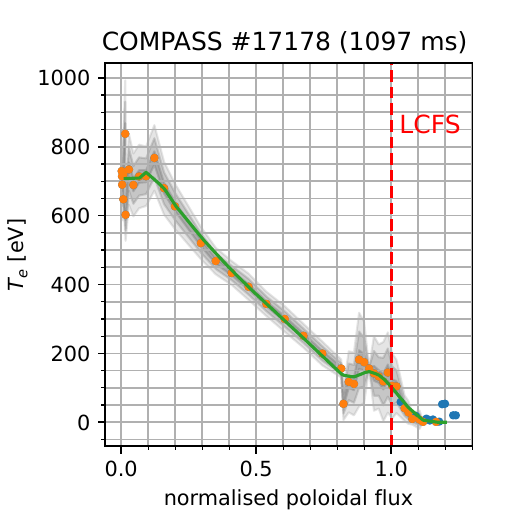}
    \caption{\label{fig:density}Electron density and temperature profile after postprocessing, derived from COMPASS Database Thomson scattering measurements.}
\end{figure}

In the case of COMPASS, a Thomson scattering diagnostic provided density and temperature profiles\cite{bilkova2018, stefanikova2016}. The profiles were extracted from the COMPASS Database and assigned to the corresponding poloidal flux surfaces. We performed outlier filtering on the data and smoothed the profiles. The measurement data are represented by dots in~Figure~\ref{fig:density}. Grey stripes shows $1 \sigma$, $3 \sigma$ and $5 \sigma$ uncertainties. The $\sigma$ standard deviation is coming from the Thomson scattering database. The orange dots represent accepted data based on our outlier filtering; the blue ones are the outliers we do not use. Green lines show the smoothed profiles. The red dashed line represents the last closed flux surface (LCFS). Smooth density and temperature profiles are the initial parameters for primary ionization modeling.

\subsubsection{Atomic physics models for primary ionization}
\label{sec:atomic}

By incorporating the excitation mechanism of the different electron shells into the ionization calculation, we achieve a model that is significantly more precise than most available ones based on numerical fitting. RENATE-OD has been validated with other codes, making it a reliable tool for neutral light alkali beam ionization modeling \cite{pokol2021}. We integrated the RENATE-OD into our synthetic diagnostic, and we provided an interface to read the plasma parameters from our preprocessing unit and export the beam attenuation profile to the numerical solver.

Looking at the literature on fusion plasma physics, the Lotz model \cite{lotz1967} is prevalent because it provides a quantitatively reliable model for a wide range of atoms. However, compared to measurements, there is disagreement around the maximum of the cross sections compared to measurements\cite{wareing1967}. Better fitting models are available \cite{uddin2005}, and two of the common ones are introduced to the synthetic diagnostic.

A more straightforward solver for electron impact ionization could be McWhirter's simplified model\cite{nrl2019, mcwhirter1965}. A more accurate though more complex model is provided by the Binary-Encounter-Bethe (BEB)\cite{kim1994}, which shows good agreement with atomic physics experiments \cite{uddin2005}.

The effect of charge exchange ionization can also be significant; therefore, further investigation has been carried out. We looked for a model that fits the measurement data well and has many applications, so we chose Tabata's models \cite{tabata1988}, which reliably describe the interactions of alkali and hydrogen atoms and ions.

\begin{figure}
    \centering
    \includegraphics[width=\mycolumnwidth]{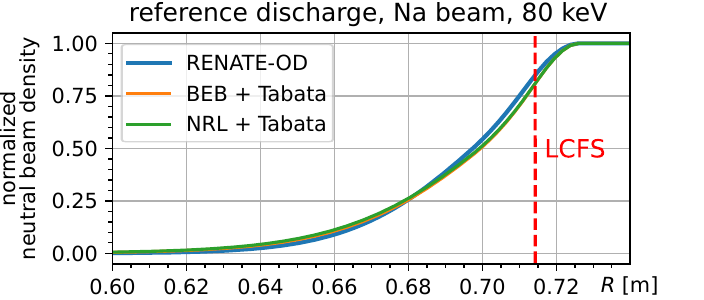}
    \caption{\label{fig:Na80}Normalized neutral beam density with RENATE-OD\cite{asztalos2019} simulation, with Binary-Encounter-Bethe \cite{kim1994} (BEB) and McWhirter's \cite{nrl2019, mcwhirter1965} (NRL) approximation with Tabata's model for direct and charge exchange ionization\cite{tabata1988} The last closed flux surface (LCFS) is indicated.}
\end{figure}

To combine the two models and integrate the cross sections, we implemented a Python code called \emph{Sibeira} (Simple Beam Ionization Rate) \footnote[3]{https://github.com/sibeira/sibeira}. We compared the modeling of RENATE-OD with the McWhirter (NRL) and Binary-Encounter-Bethe (BEB) models for different ion sources and energies. The normalized neutral beam density along the injection direction radially is shown in Figure~\ref{fig:Na80}. In the parameter range for the applications on COMPASS tokamak, we observed that the plasma parameters are outside the range of the validity of McWhirter's formula\cite{nrl2019}. Nevertheless, our calculations show a qualitatively similar ionization profile to the BEB model. The BEB model was preferred in the calculations where RENATE-OD was not available for the primary ionization.

\subsubsection{Sensitivity to magnetic perturbation}
\label{sec:perturbation}

The location of the primary ionization has been assumed to be not or only negligibly affected by the uncertainty of the exact locations of the magnetic surfaces\cite{berta2013, zoletnik2018}. This assumption was not demonstrated by numerical simulation, so it had to be made up for in the design of the synthetic diagnostics. The primary magnitude estimate of the perturbation effect is estimated using the following one-dimensional model: the magnetic flux surfaces are quasi-constant and radially stretched or compressed by a few percent. On a quasi-constant magnetic flux surface, the average kinetic energy and the $n_e V$ product can be assumed to be constant. As a first-order approximation, we used

\begin{equation}
n_e(\rho+\delta\rho) \approx \frac{n_e(\rho) }{1 + 3 \frac{\delta\rho}{\rho}}
\end{equation}

with $\frac{\delta\rho}{\rho} \ll 1$.

\begin{figure}
    \centering
    \includegraphics[width=\mycolumnwidth]{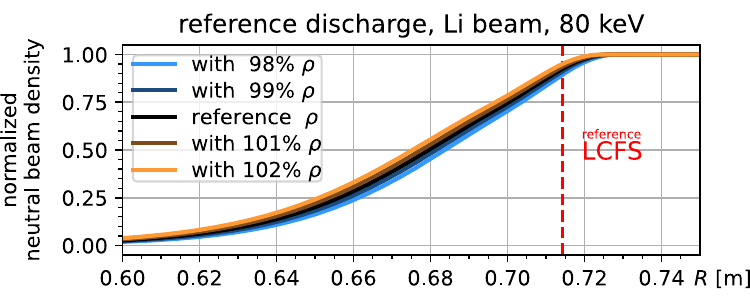}
    \caption{\label{fig:shaking}Normalized neutral beam density with different magnetic flux surface perturbations. The different colors show the influence of a virtual flux surface inflation (see legend) on the ionization profile. The last closed flux surface (LCFS) is indicated at the reference without perturbation. }
\end{figure}

A new density and temperature profile was generated from the measured data and passed to RENATE-OD as input parameters. The attenuation rates were compared in the different cases and visualized in~Figure~\ref{fig:shaking}. The different colors show different radial expansions of flux surfaces as first-order perturbations.

\begin{figure}
    \centering
    \includegraphics[width=\mycolumnwidth]{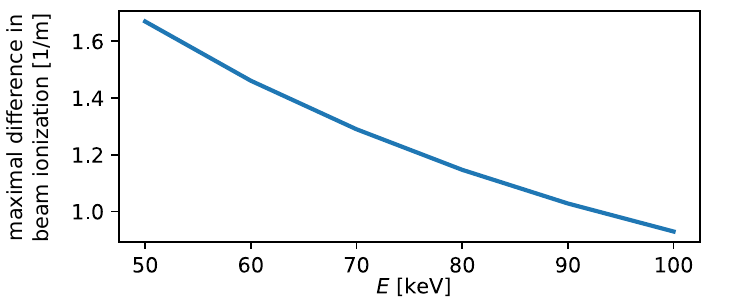}
    \caption{\label{fig:maxdiff}Maximal difference of beam ionization at different beam energy with lithium beam with $\rho'=1.01\rho$.}
\end{figure}

Some small but not insignificant differences are visible. Therefore, a wider range of beam energy was investigated and looked for the largest absolute difference in the ionization profile as shown in~Figure~\ref{fig:maxdiff}. We experienced around 10\% relative maximal difference by 1\% fluctuation in the magnetic field. The typical maximum ionization rate is 10--15 1/m, which means 7--10~cm ionization path length. The impact of the magnetic field along the neutral beam should be examined before any calculation despite previous assumptions. We can only neglect its effect if it is small compared to the uncertainty of the magnetic field and other factors presented in the article.

\subsubsection{Simplified linear beam modeling}
\label{sec:linear}

An atomic beam is three-dimensional but has a relatively narrow width compared to the dimensions of the tokamak. The maximal beam diameter is 25~mm, and we can constrict it to a smaller diameter, typically 5--10 mm\cite{zoletnik2018}. We, therefore, investigated the ionization profile using RENATE-OD at different beam thicknesses.

\begin{figure}
    \centering
    \includegraphics[width=\mycolumnwidth]{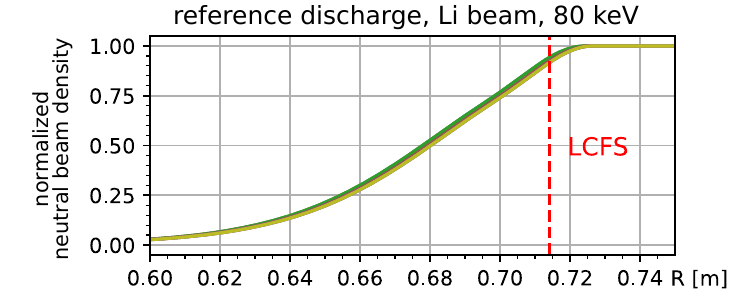}
    \caption{\label{fig:Li80_25mm}Normalized neutral beam density with maximal beam diameter (25~mm). The different colors show different distances from the midline of the beam. The last closed flux surface (LCFS) is indicated.}
\end{figure}

Figure~\ref{fig:Li80_25mm} shows the ionization profile along the beam at several points of the beam cross-section. The particle position in the cross-section has a 1--2 percent difference in the ionization coordinates. Given that the uncertainty in the magnetic field reconstruction may be larger\cite{stefanikova2016}, we can safely use a beam model for an ionization profile with zero diameters, even in this case.

The advantage of the simplified beam model is that it is unnecessary to specify an ionization rate for a three-dimensional position in the beam. However, specifying the radial position and a beam ionization profile is sufficient. The simplification significantly increases performance both for ionization modeling with RENATE-OD and for the GPU memory operations of the trajectory calculations.

\subsection{Secondary ionization loss}
\label{sec:secondary}

The Atomic Beam Probe detects singly ionized alkali ions. The main reason for using light alkali ions is that the secondary ionization energy is much higher than the primary ionization energy since the electron shells of singly ionized ions have a noble gas electron configuration. However, plasma conditions can allow a significant loss due to secondary ionization. Therefore, we have to take it into account. Since the ionization energy is high compared to the electron temperature, the estimate of the McWhirter's formula\cite{nrl2019} is within its validity  ($0.02 \lessapprox t_e \lessapprox 100$). The secondary ionization rate is \cite{nrl2019}:

\begin{equation}
     \label{eq:McWhirter}
     \frac{dn_1}{dt} \approx -  n_1 \cdot n_e \cdot \frac{\sqrt{t_e}\cdot e^{-1/t_e}}     {\left(\frac{E_{1\rightarrow 2}}{1 \mathrm{eV}}\right)^{3/2}\cdot(6 + t_e)}
     \cdot 10^{-11} \frac{\mathrm{m}^3}{\mathrm{s}} \end{equation}

where $n_1$ is the singly ionized ion beam density, $n_e$ is the local electron density in $1/\mathrm{m}^3$, $T_e$ is the local electron temperature, $E_{1\rightarrow 2}$ is the secondary ionization energy in electronvolts and $t_e = \frac{T_e}{E_{1\rightarrow 2}}$ is the normalized electron temperature.

When the concept of ABP was first conceived in 2009\cite{berta2009}, we provided an estimation for the secondary ionization for different alkali species as a source for the Atomic Beam Probe\cite{berta2009}. These calculations utilized the density and temperature profiles of the old COMPASS-D measurements from Culham\cite{berta2009}. The calculations were based on McWhirter's formula (Equation~\ref{eq:McWhirter}), besides with globally averaged density and temperature parameters\cite{berta2009}. We repeated the calculation with our new tool, using synthetic diagnostics on the reference discharge at different energies and alkali sources. The primary ionization of lithium and sodium beams is estimated with RENATE-OD. The numerical solver uses the Runge--Kutta--Nyström method. The secondary ionization yield is compared to the estimates of the ABP concept paper\cite{berta2009}. Due to the radial position of the primary ionization, the path length and trajectory of the singly ionized ions are different.
For this reason, the secondary ionization rate depends on the primary ionization's radial position. Therefore, we describe the secondary ionization loss by a mean value and standard deviation. The comparison is presented in the Table~\ref{tab:secondary}.

\begin{table}
    \centering
     \begin{tabular}{| c | r@{\% $\pm$ } r@{\%} | r@{\% $\pm$ } r@{\%} |}
         \hline
         $E_{\mathrm{beam}}$ [keV] &  \multicolumn{2}{ c |}{Li$^{+}$ $\rightarrow$ Li$^{2+}$} &         \multicolumn{2}{ c |}{Na$^{+}$ $\rightarrow$ Na$^{2+}$}\\
         \hline
 40 & 4.953 & 0.591 & 22.654 & 0.639 \\
 50 & 4.963 & 0.480 & 21.673 & 0.628 \\
 60 & 4.805 & 0.398 & 20.883 & 0.624 \\
 70 & 4.655 & 0.304 & 20.220 & 0.652 \\
 80 & 4.523 & 0.277 & 19.809 & 0.764 \\
 90 & 4.405 & 0.316 & 19.337 & 0.778 \\
 100& 4.321 & 0.291 & 19.023 & 0.833 \\
         \hline
 estimated \cite{berta2009} & \multicolumn{2}{ c |}{$\approx$ 5\%} &         \multicolumn{2}{ c |}{$\approx$ 18\%}\\
         \hline
     \end{tabular}

    \caption{Loss due to secondary ionization compared to the full beam ion population.}
     \label{tab:secondary}
 \end{table}

Qualitatively, we obtain a similar result for the secondary ionization loss as in the ABP concept publication\cite{berta2009}. The variation due to the radial position of the primary ionization shows that the calculation of secondary ionization is required during the trajectory calculation, and it impacts the detected single alkali ion population.

\section{Trajectory solver}
\label{sec:trajectory}

In this section, we elaborate on how the trajectory solver (see: Figure~\ref{fig:abpsd}) works. First, we describe the physics model of the movement of alkali ions in fusion plasmas and present the differential equation to be solved. Then, we show the different differential equation solvers and explain their benefits regarding their mathematical precision and numerical performance.

\subsection{Physical model of ion trajectories}
\label{sec:physical}

Once we know where the particles in the atomic beam are ionized, we can start calculating the trajectory of the primary ionized ions. The trajectory calculation requires solving the equation of motion for the Lorentz force numerically for each particle:

\begin{equation}
    \label{eq:newton}
 \mathbf{\ddot r} = \frac{q}{m} \left(\mathbf{E}(\mathbf{r}) + \mathbf{\mathbf{\dot r}}(\mathbf{r})\times \mathbf{B}(\mathbf{r})\right)
\end{equation}

where $\mathbf{r}$, $\mathbf{\dot r}$ and $\mathbf{\ddot r}$ are the position, velocity and acceleration of a single ion, respectively; the charge of the ion $q=e$ is the elementary charge, $m$ is the mass of an ion, $\mathbf{B}$ is the magnetic field, $\mathbf{E}$ is the electric field. Typically $|\mathbf{E}| \ll |\mathbf{\mathbf{\dot r}}\times\mathbf{B}|$ Therefore, in most of the calculations, the electric field is negligible. The electric field can have a significant influence close to the wall of the vacuum chamber, and we can use parametric modeling for it.

The trajectory of each ion can be calculated independently and in parallel. The coordinates of the particles ($\mathbf{r}$ ) are represented in a Cartesian coordinate system, where the initial toroidal coordinate of the center of the beam is zero. These coordinates are mapped to a poloidal cross-section plane in each iteration step. That is how the local magnetic and electric field is given, and the poloidal flux coordinate is determined. The loss due to the secondary ionization that is described by  Equation~\ref{eq:McWhirter} is numerically integrated with every step using the one-dimensional secondary ionization profile.

\begin{figure}
    \centering
    \includegraphics[width=0.8\mycolumnwidth]{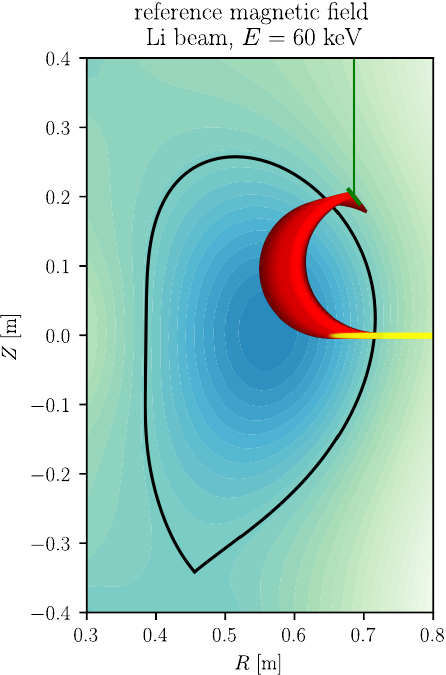}
    \caption{\label{fig:traj}Trajectories of Li$^+$ ions in the reference magnetic field.}
\end{figure}

We visualized the trajectories of all ions in~Figure~\ref{fig:traj} for one million particles with the reference discharge (COMPASS \#17178, 1097 ms, see: Figure~\ref{fig:brad}). The yellow line bar shows a 60~keV lithium atomic beam with a 5~mm diameter. The atoms are ionized along the beam, and different shades of red represent the single ionized ions. The brightness of red shows higher ion density, and the color scale is linear. The ions are detected by the ABP detector, which is green in the figure. The different azure-shaded nested ovals show the poloidal flux surfaces, and the last closed flux surface (LCFS) is drawn as a black line.

\subsection{Numerical solvers}
\label{sec:numerical}

To know the impact position of ions, we need to determine the trajectories of each ion by solving the equation of motion (Equation~\ref{eq:newton}) numerically. We explain the integration schemes that are available in the synthetic diagnostic.

First, we introduce a quadratic linearized Runge--Kutta solver \cite{numrec2007}, then the mathematically more accurate Runge--Kutta-Nyström model \cite{rkn}. The Runge--Kutta solvers are optimal for several numerical applications because of their numerical accuracy and the ideal computational complexity\cite{numrec2007}. The disadvantage is that Runge--Kutta solvers are not energy conservative and consequently not suitable for calculations with large -- several million -- numerical steps, such as transport processes\cite{qin2013}. In the case of ABP modeling, we need a small number of numerical steps -- at most a few hundred -- which makes this numerical effect negligible, as the numerical results show. To support a broader application of the synthetic diagnostic, a Boris--Verlet solver \cite{boris1970, birdsall1991, qin2013} has also been implemented as an available numerical solver.

\subsubsection{Linearized Runge--Kutta solver}
\label{sec:lrk}

The Runge--Kutta method\cite{numrec2007} is utilized to solve differential equations in the form of $\dot{y}(t) = f\left(y, t\right)$. The equation of motion (Equation~\ref{eq:newton}) includes second-order derivatives, but the derivatives are not independent of each other. Expanding the cross-products, we can rewrite the equation in the following form:

\begin{equation}
    \label{eq:lory}
    \begin{array}{l@{~=~}l@{~=~}l}
 \dot{y}_1 & v_1 = y_4 & f_1(y_4)\\[1em]
 \dot{y}_2 & v_2 = y_5 & f_2(y_5)\\[1em]
 \dot{y}_3 & v_3 = y_6 & f_3(y_6)\\[1em]
 \dot{y}_4 & \dfrac{q}{m} \left(E_1 + v_2 B_3 - v_3 B_2 \right) \\[1em]
        & \dfrac{q}{m} \left(E_1 + y_5 B_3 - y_6 B_2 \right) & f_4(y_5,y_6)\\[1em]
 \dot{y}_5 & \dfrac{q}{m} \left(E_2 + v_3 B_1 - v_1 B_3 \right) \\[1em]
        & \dfrac{q}{m} \left(E_2 + y_6 B_1 - y_4 B_3 \right) & f_5(y_4,y_6)\\[1em]
 \dot{y}_6 & \dfrac{q}{m} \left(E_3 + v_1 B_2 - v_2 B_1 \right) \\[1em]
        & \dfrac{q}{m} \left(E_3 + y_4 B_2 - y_5 B_1 \right) & f_6(y_4,y_5)\\
    \end{array}
\end{equation}

This allows us to introduce a generalized coordinate $\mathbf{y}=\left(r_1,r_2,r_3,\dot{r_1},\dot{r_2},\dot{r_3}\right)$ and a function $f$ that depends only on $\mathbf{y}$ and time and therefore the problem is given by $\dot{\mathbf{y}}=f(\mathbf{y}, t)$. The problem is solved by a fourth-order Runge--Kutta method \cite{numrec2007}.

The disadvantage of the linearized Runge--Kutta method is that the velocity is stepped in one Runge--Kutta step after the spatial coordinates, because the numerical errors accumulate. Therefore, other methods are worth exploring.

\subsubsection{Runge--Kutta--Nyström solver}
\label{sec:rkn}

The Runge--Kutta--Nyström method\cite{rkn} is an improvement for the Runge--Kutta method, and it is used to manage problems in the form of $\ddot{y}(t)= f\left(\dot{y}, y, t\right)$. The discretization scheme is the following:

\begin{equation}
    \label{eq:nystrom1}
    \begin{array}{l@{\,=\,}l}
 \mathbf{r}(t_0+\Delta t) & \mathbf{r}(t_0) + \mathbf{\dot{r}}(t_0) \Delta t + \left( \mathbf{k}_1 + \mathbf{k}_2 + \mathbf{k}_3 \right) \! \dfrac{\Delta t^2}{6} \\[1em]
 \mathbf{\dot{r}}(t_0+\Delta t) & \mathbf{\dot{r}}(t_0) + \left( \mathbf{k}_1 + 2\mathbf{k}_2 + 2\mathbf{k}_3 + \mathbf{k}_4\right) \dfrac{\Delta t}{6}
    \end{array}
\end{equation}

with the coefficients:

\begin{equation}
    \label{eq:nystrom2}
    \begin{array}{l@{~=~}l}
 \mathbf{k}_1 & \ddot{\mathbf{r}} \left( \dot{\mathbf{r}}\,,~ \mathbf{r}\,,~ t_0 \right) \\[1em]
 \mathbf{k}_2 & \ddot{\mathbf{r}} \left( \dot{\mathbf{r}} + \mathbf{k}_1 \dfrac{\Delta t}{2}\,,~ \mathbf{r} + \dot{\mathbf{r}} \dfrac{\Delta t}{2} + {\mathbf{k}_1} \dfrac{\Delta t^2}{8}\,,~ t_0 \right) \\[1em]
 \mathbf{k}_3 & \ddot{\mathbf{r}} \left( \dot{\mathbf{r}} + \mathbf{k}_2 \dfrac{\Delta t}{2}\,,~ \mathbf{r} + \dot{\mathbf{r}} \dfrac{\Delta t}{2} + {\mathbf{k}_2} \dfrac{\Delta t^2}{8}\,,~ t_0\right) \\[1em]
 \mathbf{k}_4 & \ddot{\mathbf{r}} \left( \dot{\mathbf{r}} + \mathbf{k}_3 \dfrac{\Delta t}{2}\,,~ \mathbf{r} + \dot{\mathbf{r}} \Delta t            + {\mathbf{k}_3} \dfrac{\Delta t^2}{2}\,,~ t_0 \right) \\[1em]
    \end{array}
\end{equation}

where $\ddot{\mathbf{r}}\left( \mathbf{r}, \dot{\mathbf{r}} \right)$ is defined in Equation~\ref{eq:newton}.

The solution is similar to the method given in the Subsection~\ref{sec:lrk} though more accurate due to the synchronous position--velocity co-calculation. However, this method is numerically more expensive than the linearized Runge--Kutta solver.

\subsubsection{Boris--Verlet solver}
\label{sec:bv}

The Runge--Kutta and Runge--Kutta--Nyström methods provide reasonable numerical accuracy while having a known shortcoming: they do not conserve kinetic energy. Since the trajectory path traveled by the singly ionized ions is short, about 3--5 Larmor radii, almost a semicircle,  this does not cause a significant inaccuracy. In terms of runtime, the trajectory solver would allow longer simulations, even for particles remaining in the plasma. Nevertheless, the kinetic energy loss limits these kind of applications.

This shortcoming can be overcome by adding a numerical solver that is kinetic energy conservative. The Boris method \cite{boris1970, birdsall1991, qin2013} was chosen for it because of its widespread use, reliability, and simplicity. It is a second-order Verlet solver with a Boris push in the velocity space\cite{boris1970}. We can introduce vectors:

\begin{figure}
    \centering
    \includegraphics[width=0.5\mycolumnwidth]{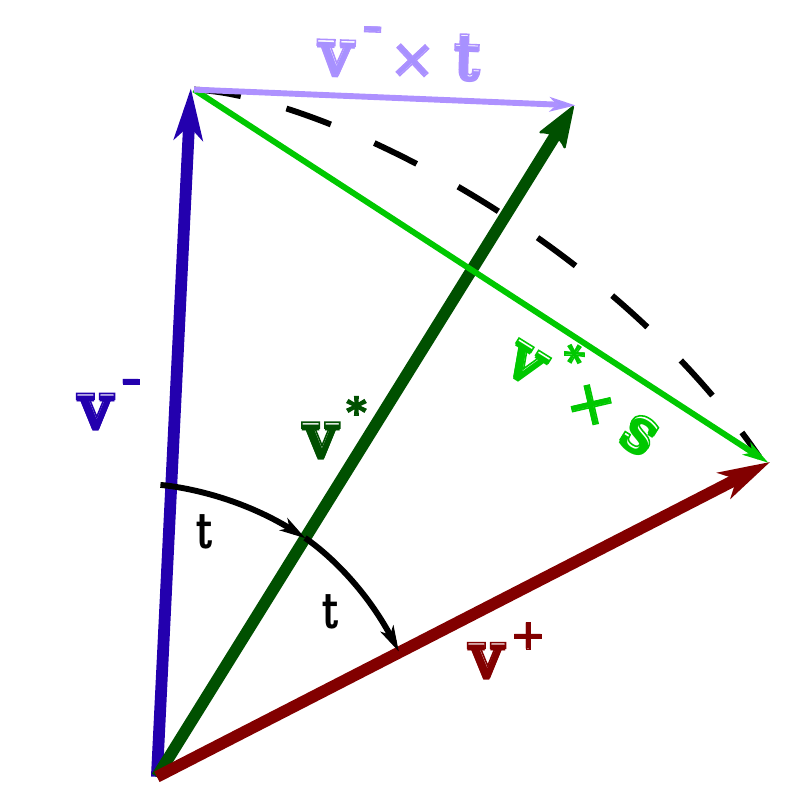}
    \caption{\label{fig:boris}Boris push in velocity space \cite{boris1970}.}
\end{figure}

\begin{equation}
    \label{eq:boris}
    \begin{array}{l@{~=~}l}
 \mathbf{v}^- & \dot{\mathbf{r}}(t_0) + \dfrac{q}{m} \cdot \mathbf{E} \cdot \dfrac{\Delta t}{2} \\[1em]
 \mathbf{v}^* & \mathbf{v}^- + \mathbf{v}^- \times \mathbf{t} \\[1em]
 \mathbf{v}^+ & \mathbf{v}^- + \mathbf{v}^* \times \mathbf{s}
    \end{array}
\end{equation}

where $\mathbf{t} = \dfrac{q}{m} \cdot \mathbf{B} \cdot \dfrac{\Delta t}{2}$ and $\mathbf{s} = \dfrac{2\cdot\mathbf{t}}{1+\mathbf{t}\cdot\mathbf{t}}$,

as it is presented in Figure~\ref{fig:boris}\cite{boris1970, birdsall1991, qin2013}. This small correction ensures the conservation of kinetic energy at every time step.

Combining the equation of motion (Equation \ref{eq:newton}) with the Boris push (Equation~\ref{eq:boris}), the coordinates can be written as:

\begin{equation}
    \label{eq:boris2}
    \begin{array}{l@{~=~}l}
 \dot{\mathbf{r}}(t_0+\Delta t) & \mathbf{v}^+ + \dfrac{q}{m} \cdot \mathbf{E} \cdot \dfrac{\Delta t}{2} \\[1em]
 \mathbf{r}(t_0+\Delta t) & \mathbf{r}(t_0) + \dot{\mathbf{r}}(t_0) \cdot \Delta t
    \end{array}
\end{equation}

\subsubsection{Solver comparison and verification}
\label{sec:solvers}

\begin{figure}
    \centering
    \includegraphics[width=\mycolumnwidth]{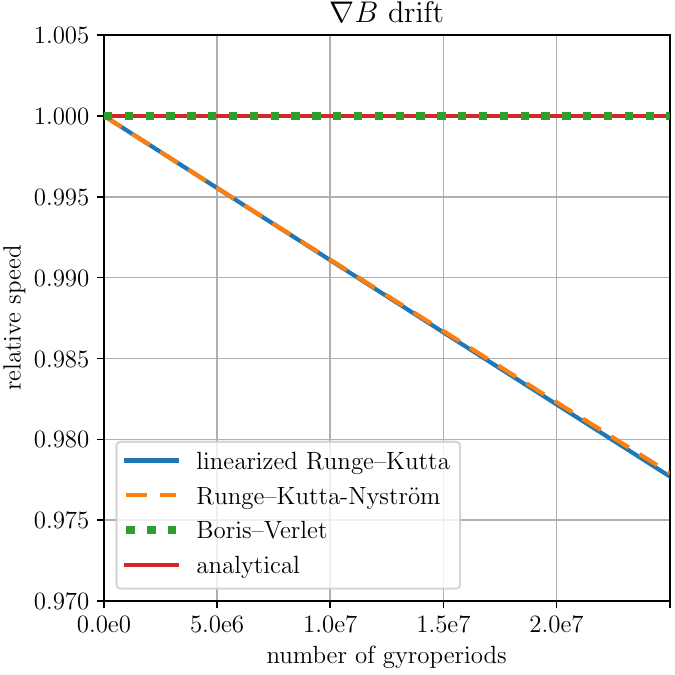}
    \caption{\label{fig:gradb_speed}Relative velocity loss for different solvers in $\nabla B$ magnetic field with $10^{-9}$~s time step.} 
\end{figure}

\begin{figure}
    \centering
    \includegraphics[width=\mycolumnwidth]{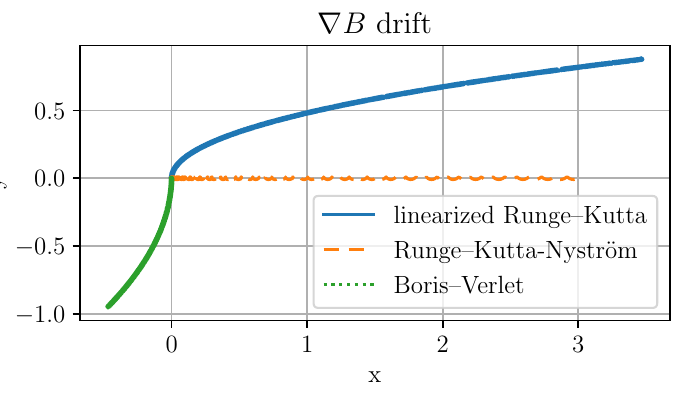}
    \caption{\label{fig:gradb_coord}Ion trajectories for different solvers in $\nabla B$ magnetic field in analytical guiding center centered coordinate system with $10^{-9}$~s time step for 5 million gyroperiods.}
\end{figure}

\begin{figure}
    \centering
    \includegraphics[width=\mycolumnwidth]{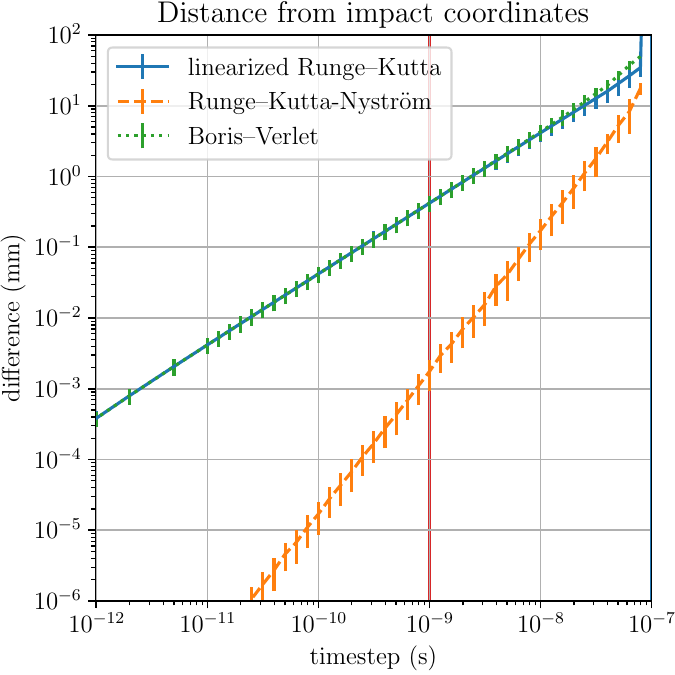}
    \caption{\label{fig:solvers}Mean and standard deviation from the impact coordinates for a three-dimensional beam with different numerical solvers and time steps.}
\end{figure}

The numerical solver is the heart of the trajectory calculator; its reliability is vital for the synthetic diagnostic. For each of the three numerical solvers, numerical tests were performed to compare the computational results for analytically well-described cases\cite{qin2013, shali2023}, and the results are presented in Figures~\ref{fig:gradb_speed}, \ref{fig:gradb_coord} and \ref{fig:solvers}. The following scenarios were used:

\begin{center}
    \begin{tabular}{ c c c }
 description & $\mathbf{B}$ & $\mathbf{E}$ \\
    \hline
 homogeneous magnetic field & $(0,0,1)$ & $(0,0,0)$\\
 homogeneous electric field & $(0,0,0)$ & $(0,0,1)$\\
    $\nabla B$ magnetic field & $(0,0,1+0.01x)$ & $(0,0,0)$\\
    $\sim 1/R$ magnetic field & $\left(0,0,\dfrac{1}{\sqrt{x^2+y^2}}\right)$ &  $(0,0,0)$\\
 magnetic field dominant & $(0,0,1)$ & $(0,0,0.1)$\\
    \end{tabular}
\end{center}

Figure~\ref{fig:gradb_speed} shows the relative loss of velocity as a function of gyroperiod due to the numerical lack of conservation of kinetic energy with $10^{-9}$~s time step. We can see that 1\% of the velocity is lost every 2.2 million gyroperiods for Runge--Kutta methods. Boris--Verlet conserves energy, as it was concerned above. The figure shows for $\nabla B$ drift, but the same figure was received for all analytical energy-conservative cases.

Deviation from the analytical trajectory can be caused not only by kinetic energy loss but also by numerical error. Therefore, we examined this impact on the scenarios mentioned in the table and visualized the trajectories in the analytical guiding center-centered coordinate system in Figure~\ref{fig:gradb_coord}. As before, the effect seemed to be negligible for a small number of gyroperiods. The Boris--Verlet seemed to shift the least compared to the analytically calculated value. On the other hand, while the linearized Runge--Kutta and Boris--Verlet move off the $\nabla B$ axis, the Runge--Kutta--Nyström remains on the axis.

We then compared the numerical solvers on the reference discharge data with experimentally relevant parameter range: a 25~mm diameter lithium atomic beam with an energy of 70~keV was injected in the simulation. The 4~cm region near the last closed flux surface (LCFS) was investigated. We scanned a wide range of numerical time steps from $10^{-8}$ to $10^{-13}$ seconds with different numerical solvers and compared the ion impact coordinates at the chosen time step to $10^{-13}$. We visualized them in Figure~\ref{fig:solvers}. We also compared the results of different solvers for a time step of $10^{-13}$ and found them to be nearly identical, with a difference only in the eighth decimal position for each ion. 

The figure shows that the Runge--Kutta--Nyström solver seems to be the most reliable and should be used.
We then achieve a numerical accuracy of less than 2 micrometers, an acceptable error compared to other uncertainties.

\subsection{Numerical background of the high-performance solver}
\label{sec:hpc}

Due to our work with a large number -- typically $10^6$--$10^7$ -- of independent particles, we could parallelize our code and, therefore, an obvious choice was to use GPGPU (general-purpose computing on GPUs) solutions\cite{cuda_example}. We chose CUDA C language by virtue of wide support\cite{cuda_example}, and we followed the principles of CUDA Performance Guidelines\cite{cuda_guidelines}. We used SDK version 2.1 to support old (Fermi) graphic cards.

First, the spline coefficients and primary and secondary ionization profiles are read from the preprocessing database and copied to the global GPU memory generated by the magnetic field submodule (see:~Section~\ref{sec:magnetic}). That is the \emph{init} box in the Figure~\ref{fig:abpsd}. Subsequent steps are performed on the GPU and are represented by green boxes.

The trajectory solver has two initialization modes: stochastic and deterministic. In stochastic initialization, we use the primary attenuation profile as a cumulative distribution function and generate the initial particle position by randomly sampling it. In the deterministic initialization, we define a grid with representative ions along the beam, and each virtual ion gets a weight according to the ionization density function. The advantage of both methods is that little data needs to be copied from the CPU to the GPU, which is runtime intensive as the initial coordinates are generated on the GPU.

The ion trajectories are solved parallelly in the computation, effectively increasing the trajectory solver's performance. At the end of each computation, the impact coordinates are determined for each ion. Then, the code synchronizes threads to ensure that all the computation is finished. If the ion is trapped or the particle cannot reach the detector plane -- due to anomaly or maldefined plane -- an error value is assigned to the ion. The synthetic data processing unit evaluates the ion trajectories -- typically millions of spatial coordinates -- to a more compact format and returns the post-processed dataset to the CPU.

\section{Synthetic data processing}
\label{sec:synthetic}

The trajectory module explained in the previous section describes the motion of the singly ionized alkali ions in the plasma. Subsequently, the detection of the ions needs to be modeled in order to obtain a complete synthetic diagnostic signal.

\subsection{Detector unit}
\label{sec:detector}

The \emph{Trajectory simulator of ABP Ions with GPU Acceleration} (briefly \emph{TAIGA}) trajectory solver iteratively computes the evolution for each alkali ion trajectory. A computation stops when a particle passes through the detector plane or reaches the maximum number of iterations. In the latter case, an error value is returned. In the former case, an interpolation is applied to the detector plane using coordinates from the previous iteration step.

\subsubsection{Impact coordinates}
\label{sec:impact}

In the initialization phase of the calculation, the normal vector of the detector plane and the geometry of the segments are loaded into the global GPU memory. We assumed a detector geometry with optionally non-uniform segment size and gap width on a perpendicular grid, which describes the detector geometry at COMPASS tokamak. The impact coordinates are mapped to the detector plane with a coordinate transformation.

The solver only stores the position and velocity coordinates of the current and previous iteration steps to optimize the numerical performance. 
We can use linear interpolation or quadratic Bezier interpolation to get the impact coordinates. The difference is negligible if we choose the right time step value. We usually use the former because it is less computationally intensive.

\begin{figure}
    \centering
    \includegraphics[width=\mycolumnwidth]{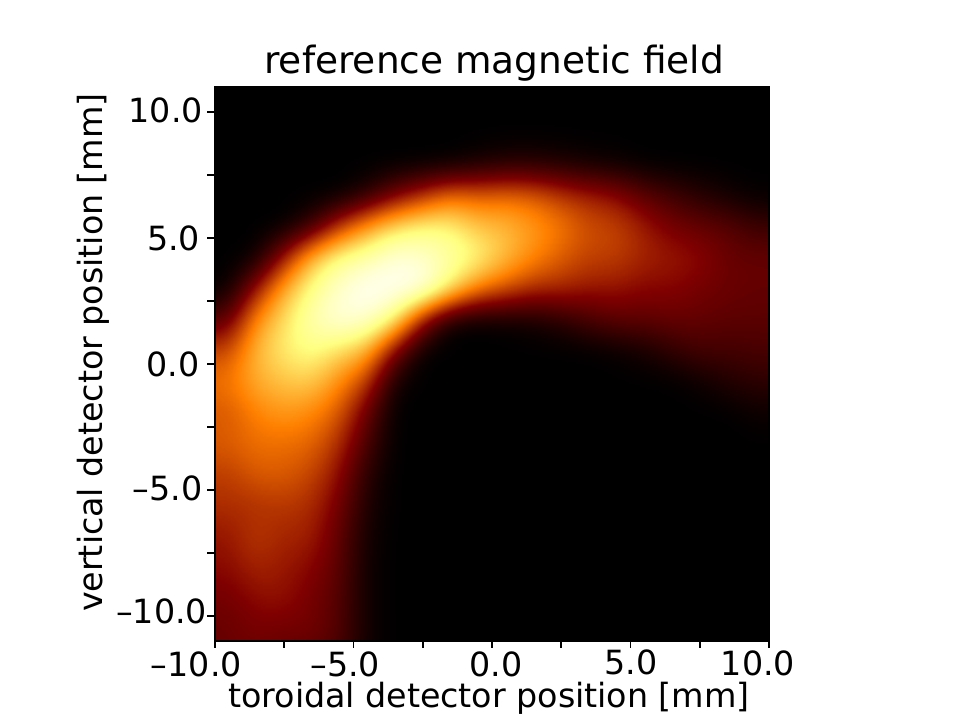}
    \caption{\label{fig:detplane}Distribution of a singly ionized 60~keV Li$^+$ beam on the detector plane from the synthetic diagnostic. The relative ion density is presented by brightness on a linear scale.}
\end{figure}

Figure~\ref{fig:detplane} represents the relative ion density on the intersection of a 60~keV single ionized lithium ion beam and the detector plane\cite{zoletnik2018} that is used in the reference discharge. Brighter colors show a higher density; the color scale is linear.

\subsubsection{Discretization}
\label{sec:discrete}

The detector postprocessor unit assigns the interpolated ion position to the right detector segment in the detector matrix. Each ion has a segment ID value or an error indicator showing if the particle is undetected. To count the ions on each segment, we need an impact counter postprocessor submodule, which atomically sums each of the ions to a segment counter to ensure another thread does not overwrite the counter. At the end of the postprocessing, we need to divide the counter values by the total number of virtual ions and multiply by the injection beam current to get the electric current on each detector segment weighted by the loss due to secondary ionization.

One option is to return with the impact coordinates and evaluate them with an external processing unit.
The other option is that instead of returning the impact coordinates of millions of particles, the evaluation can be processed on the graphics card. With this step, we typically transfer a two-digit element array of the counters corresponding to the detector segments. This discretization step is parallelized as well, which provides an effective evaluation compared to on-CPU postprocessing. The decreased return value size means a significant performance improvement due to memory copy bandwidth loss by avoiding the generation of megabytes of data, which will only be a few hundred bytes.

\begin{figure}
    \centering
    \includegraphics[width=\mycolumnwidth]{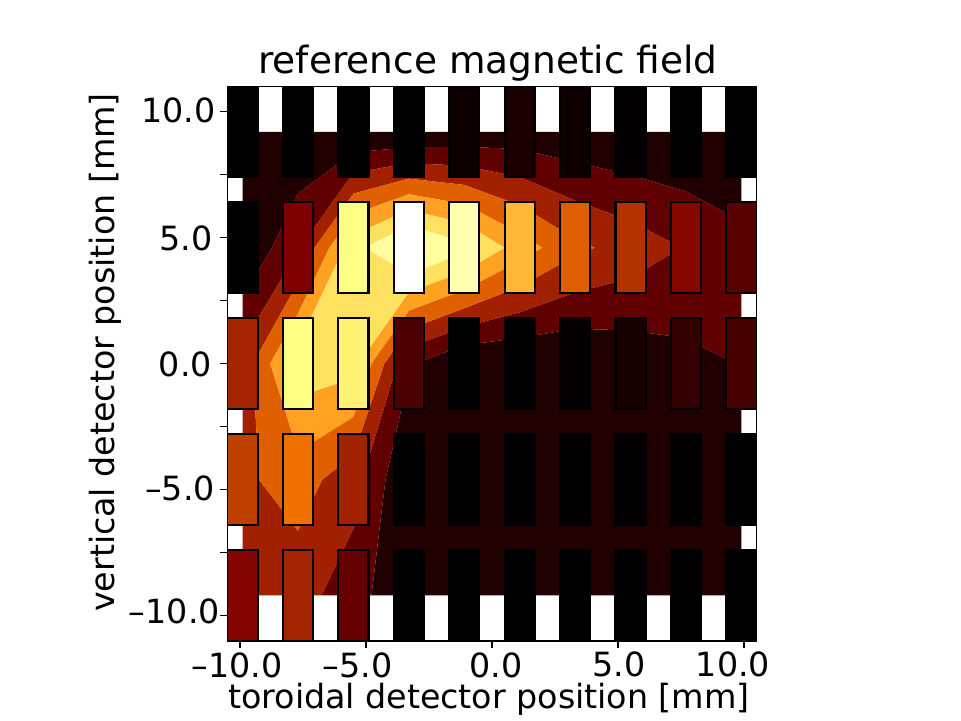}
    \caption{\label{fig:detsegm}Relative current of Li$^+$ ions on each detector segment from the synthetic diagnostic. In the background, a graphical interpolation is presented to provide a better visual comparison and understanding.}
\end{figure}

Figure~\ref{fig:detsegm} shows the discretization of the singly ionized lithium beam shown in Figure~\ref{fig:detplane}. The detector segments follow the detector geometry used in the real measurement\cite{zoletnik2018, refy2019}. Rectangles with black borders illustrate each segment. The fill color of the rectangles indicates the relative ion current: light shades imply a higher current, and the color scale is linear. The currents of the surrounding detector segments were interpolated and visualized in the gaps between the detector segments to make the two figures easier to compare visually.

The secondary electrons produced by the alkali ions significantly increased the detected current of the Faraday cup cells~\cite{hacek2018, refy2019}. For this reason, a double-biased mask was placed in front of the detector to suppress the secondary electrons in the experiments. The synthetic diagnostic, therefore, does not include a module for secondary electrons. However, if it is needed in the future,  a secondary electron module can be added after the discretization. The secondary electron trajectory and the additional detector current can be modeled by the TAIGA trajectory solver module, with the electron as species and the detector plane coordinates as initial coordinates.

\section{Applications}
\label{sec:app}

The synthetic diagnostic and the GPGPU trajectory simulator core module have provided valuable support during the diagnostic design, scenario planning, and operation phases of the ABP measurements. The subsequent section presents instances of these steps and provides a projection of the potential further application opportunities.

\subsection{Support of scenario development}
\label{sec:scenarios}

The development of the numerical solver accompanied the design of the COMPASS ABP detector, such as the prediction of the ion trajectories, the aiding the positioning of the detector, and tuning the grid resolution for different alkali beams with different beam energies \cite{berta2013, hacek2018, zoletnik2018}. Examples of results obtained by the trajectory calculator, aiding the diagnostic planning, were published\cite{berta2013, hacek2018, zoletnik2018}.

However, applying the \emph{TAIGA} trajectory simulator did not stop at the design of the diagnostic. Support was provided by outlining different scenarios before and during the measurement campaigns. The goal was to detect the alkali ion beam always in the optimal position. The input parameters were investigated in different typical magnetic field configurations, beam materials, beam energies, deflection angles, and detector positions since we could move the detector vertically and slightly perpendicularly to the poloidal plane.

\begin{figure*}
    \begin{minipage}{0.3\textwidth}
        \includegraphics[width=\linewidth]{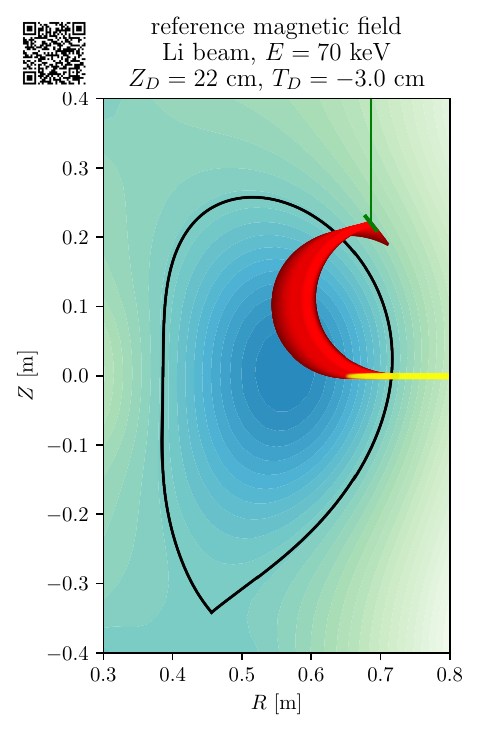}
        \centering ~\hspace*{1cm}(a)
    \end{minipage}
    \begin{minipage}{0.3\textwidth}
        \includegraphics[width=\linewidth]{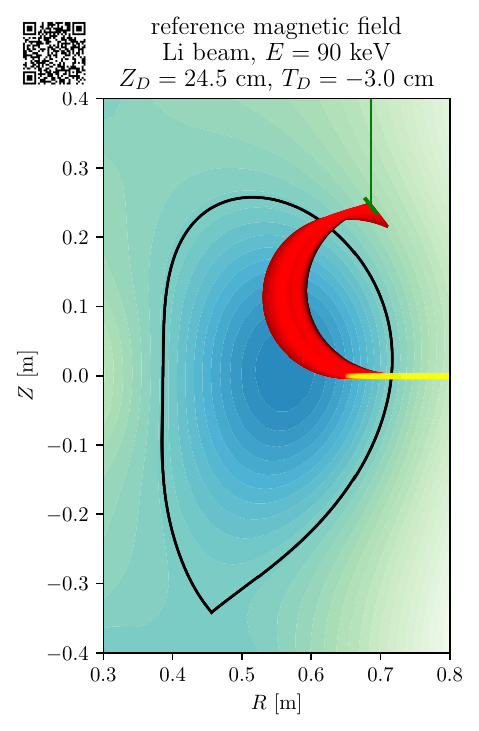}
        \centering ~\hspace*{1cm}(b)
    \end{minipage}
    \begin{minipage}{0.3\textwidth}
        \includegraphics[width=\linewidth]{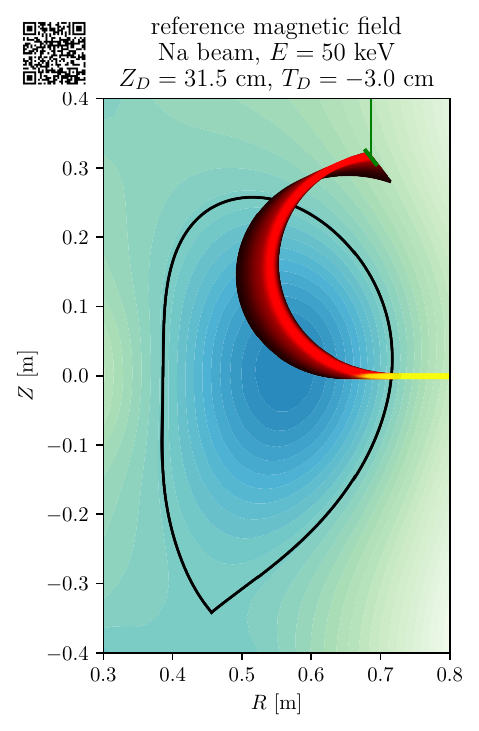}
        \centering ~\hspace*{1cm}(c)
    \end{minipage}
    \caption{Trajectories of 70 (a) and 90 keV lithium (b), and a 50 keV sodium (c) beam at reference discharge.}
    \label{fig:trajref}
\end{figure*}

In the following, we show results for two discharges. The first one, referred \emph{reference discharge}, has been introduced in the preceding part of the paper in Section~\ref{sec:init}: COMPASS~\#17178 at~$t=1097$~ms, and we are referring to reference discharge. The plasma current was $I_p=159.2$~kA, which is not a peak operation of COMPASS but optimal for the ABP. In the simulation, a 5~mm diameter atomic beam was used with a perpendicular injection to the plasma, a lithium beam with 70~keV energy at Figure~\ref{fig:trajref}a, 90~keV energy at Figure~\ref{fig:trajref}b and a 50~keV sodium beam at Figure~\ref{fig:trajref}c. We present the most optimal detector position cases of simulation scenarios, where $Z_D$ shows the vertical position and $T_D$ is the toroidal Cartesian shift of the geometrical center of the detector from the injection coordinates. The yellow line is the atomic beam, and the different shades of red present the ion beam, where toned color means a larger number of ions.

Comparing the simulation results, we can determine how the vertical position of the detected ion depends on the beam energy. For example, at the reference discharge with lithium ion, it is around $dZ_D/dE_{beam}\approx1.2-1.4$~mm/keV, but it also depends on the magnetic configuration. What is physically more interesting is that the sodium beam ionizes mainly around the last closed flux surface. The evolution of this plasma region obtains relevant information about the evolution of plasma edge instabilities, which was the initiative for the ABP concept.

\begin{figure*}
    \begin{minipage}{0.3\textwidth}
        \includegraphics[width=\linewidth]{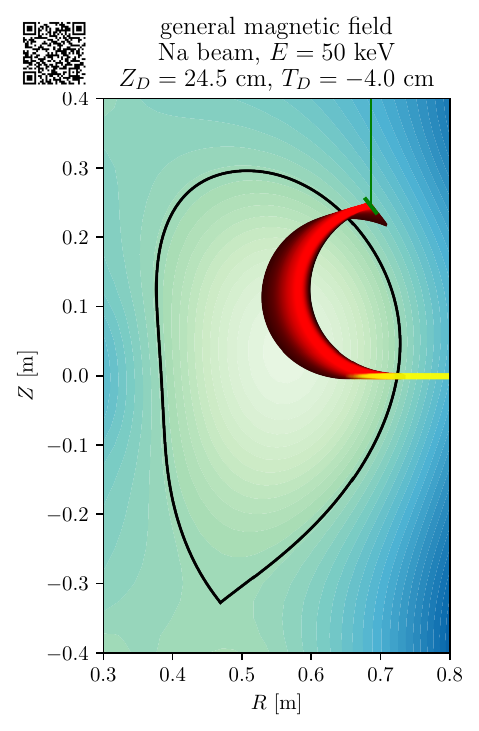}
        \centering ~\hspace*{1cm}(a)
    \end{minipage}
    \begin{minipage}{0.3\textwidth}
        \includegraphics[width=\linewidth]{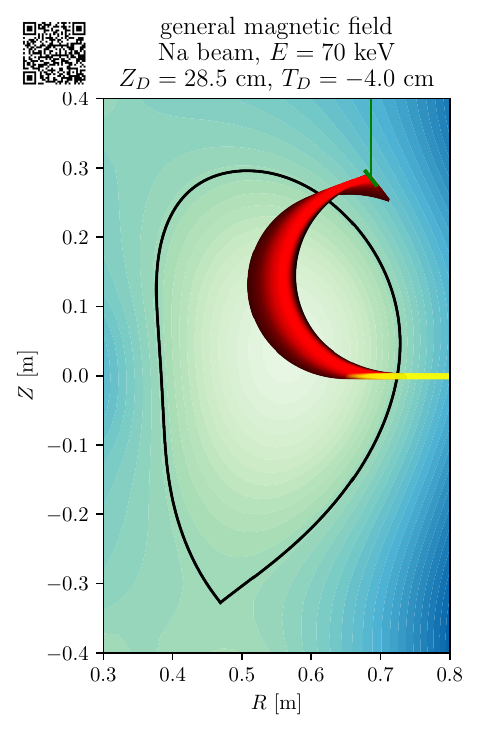}
        \centering ~\hspace*{1cm}(b)
    \end{minipage}
    \begin{minipage}{0.3\textwidth}
        \includegraphics[width=\linewidth]{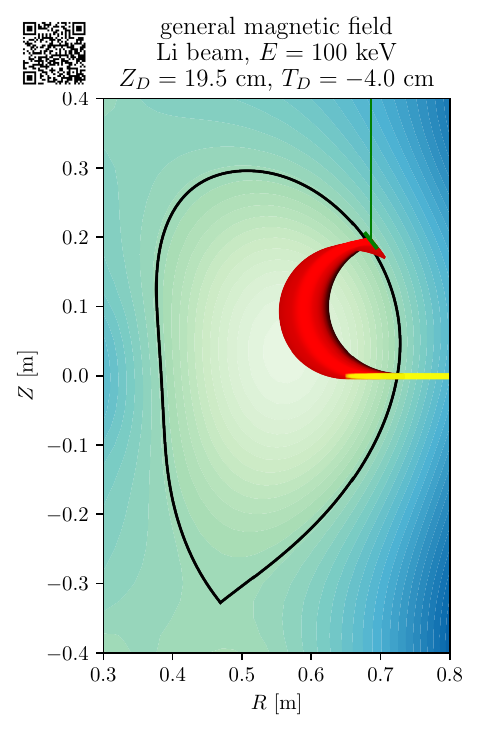}
        \centering ~\hspace*{1cm}(c)
    \end{minipage}
    \caption{Trajectories of 50 (a) and 70 keV sodium (b), and a 100 keV lithium (c) beam at general discharge.}
    \label{fig:trajgen}
\end{figure*}

The aforementioned magnetic field configuration required a dedicated discharge scenario, which limited the ABP measurements. Therefore, the experiment team investigated the possibility of operating the ABP in a different configuration. The COMPASS \#19655 discharge at~$t=1175$~ms is a \emph{general discharge} which are presented in Figures~\ref{fig:trajgen}a--c. It has a stronger magnetic field, and the plasma current was $I_P=250.3$~kA, being a typical COMPASS discharge scenario. Many measurements were performed in this magnetic scenario and similar ones. Therefore, the general discharge was numerically evaluated during and after the measurements by the TAIGA-SD synthetic diagnostic. Similarly to the reference discharge, a 5~mm alkali atomic beam was perpendicularly injected, though with a sodium beam and higher beam energies.

We repeated trajectory simulations with 50--100~keV lithium and sodium beams with 5~mm diameter and perpendicular injection. Detection would not be possible with a sodium beam of lower energy or lithium beam of any energy of the operation since, as seen in Figure~\ref{fig:trajgen}c,  the detector would have to be positioned proximal to the last closed flux surface. Furthermore, we found a uniform ionization distribution along the entire radial profile, which is undesirable. In contrast, sodium beam simulations show optimal profiles and detection geometries. We found that the beam attenuation caused by secondary ionization is around 5\%, which is still tolerable.

\subsection{Comparison with measurements}
\label{sec:measurement}

\begin{figure*}
    \centering
    \includegraphics[width=0.45\textwidth]{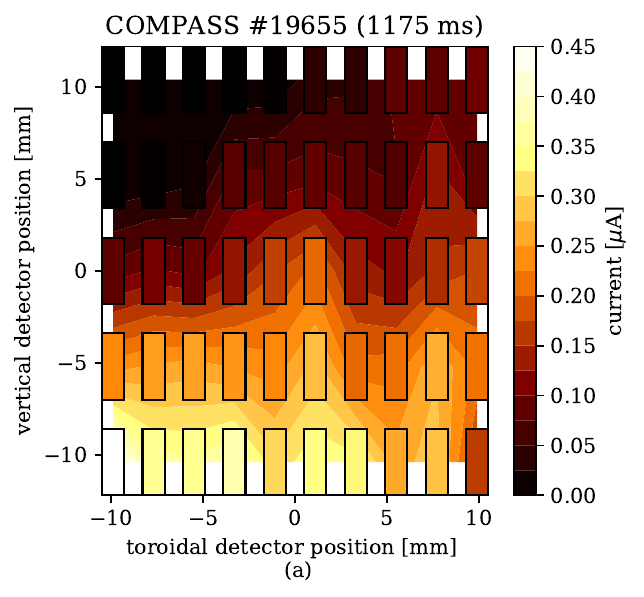}
    \includegraphics[width=0.45\textwidth]{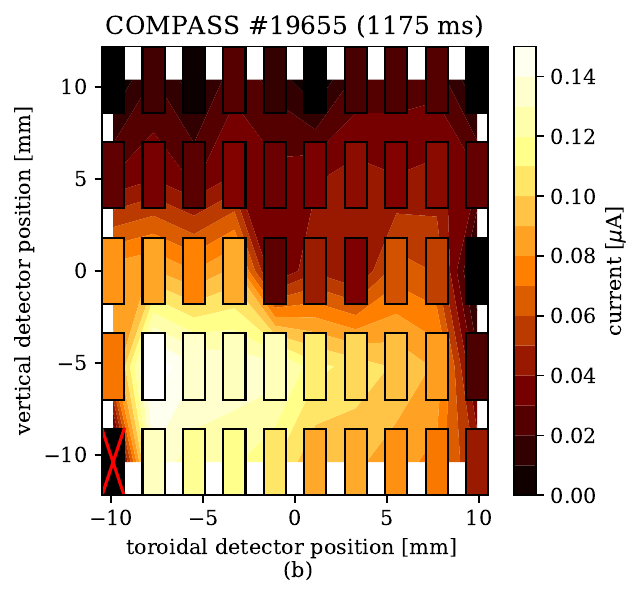}
    \caption{\label{fig:19655}Comparison of the synthetic signal of the Atomic Beam Probe (a) with measurement data\cite{refy2019, refy2021} (b).}
\end{figure*}

During the last period of the COMPASS tokamak's operation, a number of campaigns took place. The beam was observed with good certainty in the location predicted by the synthetic diagnostic, and it responded to changes in the large timescale magnetic field similarly to what we expected in many cases.

During the \#19655 discharge, good-quality measurement signals were collected from ABP, which we could use for a qualitative comparison. A 66~keV sodium beam with a reduced 5~mm diameter was injected with 1-kilohertz modulation\cite{refy2021}. The input parameters of the simulation were set from the real experiment, such as the beam deflection, the detector position, and its tilt angle.

We read the magnetic flux data from the EFIT++ database, and the magnetic field submodule of the synthetic diagnostic restored the magnetic field at different time steps of the \#19655 discharge. One-dimensional electron density and temperature profiles from Thomson scattering measurements were extracted from the COMPASS Database. The two-dimensional spline coefficients were generated, stored in our database, and passed to the trajectory solver submodule as an initial parameter. The ionization submodule scanned the density and temperature profiles and provided the input to the RENATE-OD\cite{guszejnov2012, asztalos2019} rate coefficient solver. The ionization submodule retrieved the radial ionization profile from the output of the RENATE-OD and passed it to the trajectory solver submodule as an initial parameter. The coordinate initialization submodule determined each virtual ion's initial spacial and velocity coordinates on the graphics processor for the trajectory calculation. The TAIGA trajectory solver submodule solved the equation of motion for all the one million sampled ions. With such a large number of particles, the stochastic uncertainty of the calculation, based on repeated runs, was negligible. The trajectory calculation was computed on a separate graphical thread for each ion. The calculation was interrupted automatically when each ion passed the detector plane. The postprocessing submodule summed the number of virtual ions per detector segment, considering the loss due to secondary ionization and returning with the relative ion current. The synthetic diagnostic multiplied them by the estimated injected particle current (particle/second).

We compared the detector signal from the measurement with the synthetic diagnostics signal, as shown in Figure~\ref{fig:19655}. The figure on the left (a) shows the signal of the synthetic diagnostic, and the figure on the right (b) is the published measurement result\cite{refy2021}. Despite minor differences, we have seen that the detection location and the size of the beam overlap well with the simulation results. We have overestimated the amplitude compared to the measurement, possibly due to the loss of the injected atomic beam or detection loss. Further calculations and modeling are needed for clarification.

\subsection{Further use cases}
\label{sec:outlook}

The experience with measurements and modeling synthetic diagnostics shows positive perspectives; nonetheless, further studies are needed to investigate the diagnostic's applicability according to the sensitivity of magnetic field perturbations and its possible installation in other devices.

In the future, we will investigate numerous cases covering an extensive range of spatial distribution of magnetic field perturbations. The synthetic diagnostics' high-performance GPGPU solver allows us to simulate a large number of cases efficiently. By looking at these simulations, we aim to gain a stronger understanding of the sensitivity of atomic beam probes and help the fusion plasma diagnostic community design and operate similar diagnostics in the future.

\section{Conclusion}
\label{sec:conclusion}

In this article, we presented the concept of synthetic diagnostics for \emph{Atomic Beam Probe (ABP}). The toolset is named (\emph{TAIGA-SD}), which includes the \emph{Trajectory simulator of ABP Ions with GPU Acceleration (TAIGA)} trajectory solver.

The Atomic Beam Probe is a novel diagnostic tool installed on the COMPASS tokamak to detect fast magnetic field changes in fusion plasmas. The diagnostic relies on a neutral atomic beam being injected into the plasma on the midplane. The alkali atoms are singly ionized in the plasma, and the magnetic field then governs their trajectories. These particles leave the plasma at the top of the confined region. At the COMPASS tokamak, a Faraday cup matrix detector captured and detected the alkali ions.

As the first step of synthetic diagnostic development, we determined the physical processes relevant to the synthetic signal. We designed the synthetic diagnostics and defined its components. We went through each component in the recent paper, trying to identify which physical processes can be dominant and which are negligible compared to the other processes. We supported it with calculations and modeling.

The synthetic diagnostic was built so that the individual units would be easily replaceable if required in the future.  At the core of the code is a high-performance GPU trajectory solver, surrounded by an atomic physics input and a postprocessing unit. The ionization of neutral alkalis to singly ionized ions is modeled by RENATE Open Diagnostics\cite{guszejnov2012, asztalos2019}, a collisional--radiative model for neutral hydrogen isotope and alkali beams. We added our combined Binary-Encounter-Bethe--Tabata model for the beam species where the RENATE-OD is unavailable. The trajectory solver is the result of a long-time elaboration supporting detector design and development \cite{hacek2018, zoletnik2018}. We have identified some shortcomings of the numerical solver and added better numerical solvers, which were presented in detail. The synthetic data processing section presented the conversion of the particle trajectories to physical signals.

The output of the synthetic diagnostics gives reasonable indications with the experiments, and the synthetic signal follows the change of the magnetic field. However, to design future diagnostics, further feasibility studies may be required to determine what magnetic field variations can be detected and to deduce the limitations of the Atomic Beam Probe. The here presented \emph{TAIGA-SD} synthetic diagnostic could serve this purpose, where a high-performance computing core can significantly speed up the computational runtime for the simulations.

\pagebreak
\section*{Acknowledgments}
This work has been carried out within the framework of the EUROfusion Consortium, funded by the European Union via the Euratom Research and Training Programme (Grant Agreement No 101052200 -- EUROfusion). Views and opinions expressed are however those of the author(s) only and do not necessarily reflect those of the European Union or the European Commission. Neither the European Union nor the European Commission can be held responsible for them.

This work was co-funded by MEYS project LM2023045.

The first author received an FI scholarship supported by the Ministry of Research and Universities of the Government of Catalonia and the European Social Fund related to the recent paper.

\pagebreak
\bibliographystyle{style/ans_js}
\bibliography{abp}

\end{document}